# Lossless Secret Image Sharing Schemes

**Binu V. P[1] and A. Sreekumar[2]**

[1,2]Department Of Computer Applications, CUSAT

Cochin-22, kerala, India

[1]binuvp@gmail.com, [2]sreekumar@cusat.ac.in

**Abstract**

Secret image sharing deals with splitting confidential images into several shares and the original image can be reconstructed from the qualified subset of the shares. Secret sharing schemes are used in transmission and storage of private medical images and military secrets. Increased confidentiality and availability are the major achievements. We propose an efficient (2, 2) scheme and (2, 3) scheme for secret image sharing. The scheme is lossless and also the share size is same as the secret size. The sharing and revealing phase uses simple modular arithmetic which can be very easily implemented. Experimental results on Binary and Gray scale images show that the proposed scheme is secure and efficient.

## 1. Introduction

The storage and exchange of secret data or images is a big concern. Encryption makes data unintelligible to read but the chance that the secret may be lost or manipulated. This is because there is only a single point of failure. If we use many duplicates to overcome the weakness, the danger of security exposure will also increase. Secret sharing might be one of the possible solutions. In secret sharing the secret is divided into several pieces called shares such that the shares will not explore any partial information of the secret. The shares can later be combined to reveal the secret. In the threshold scheme a qualified subset of shares are needed to recover the secret. An unqualified subset can gain absolutely no information about the secret are called perfect schemes.

Blakley [3] and Shamir [2] independently proposed the concept of secret sharing. The scheme introduced was $(t,n)$ threshold scheme. A $(t,n)$ threshold secret sharing scheme is a cryptographic primitive used to distribute secret $s$ into $n$ participants in such a way that a set of $t$ (or more) participants can recover the secret. Their scheme was highlighting the scenario of safeguarding the cryptographic keys by sharing it with $n$ participants and any $t$ of them can join together to reveal the secret key. These schemes cannot be applied directly to images. It is difficult to consider the image as a single secret. If we treat each pixel value as secret then a large number of polynomials have to be manipulated to share and reveal the secret with Shamir's scheme. Thien and Lin [4] proposed a secret image sharing scheme based on Shamir's secret sharing scheme. They used a $(r,n)$ scheme where $r$ sub pixel values are used as coefficient of the polynomial to generate shares. Large number of polynomials has to be manipulated to generate and reveal the secret and also the basic method is lossy. Tso [7] used Blakley's concept for secret image sharing. Geometry based approach is proposed by Chen et al [8].

Naor and Shamir [6] invented a new type of secret sharing scheme called Visual Cryptography (VC) scheme. The scheme can be used to decode secret from shares directly without performing any computation. Recovery scheme employed by the VC is based on Human Visual System. The scheme is more suitable for binary images. Pixel expansion and limited contrast are major issues.

Number theoretic features are used in the field of secret sharing. Concept of Chinese Remainder Theorem and Quadratic Residues has been applied in different studies. Chen and Chang [5] use quadratic residue technique for secret image sharing. They proposed a *(2, 2)* scheme which is lossy and the lossless scheme having the share size larger than the secret. The computations involved is also more. Guzin et al [10] modified the scheme on quadratic residues to make it $(t,n)$ by combining it with Thein scheme.

In this paper we propose a *(2, 2)* scheme for secret image sharing where both the shares are needed to reconstruct the original secret image. However the scheme is very efficient compared with the recent schemes proposed by Chen and Chang [5] using quadratic residues. Their scheme is lossy and in the lossless scheme one of the share size is larger than the secret image. Our scheme uses simple number theoretic concepts and is very efficient since the computations involved in making the shares and reconstruction phase is of less complexity. We also introduced a *(2,3)* scheme where a secret image can be split into three shares and any two valid shares can be used to re construct the original secret image. The proposed *(2,3)* scheme can be efficiently implemented compared with the implementation of Shamir's scheme for secret image sharing by Thein [4].

In the rest of this paper, Section 2 describes mathematical background. Section 3 describes the proposed secret image sharing scheme. Security analysis and Experimental results are given in section 4 and section 5. Conclusions is stated in section 6.

## 2. Mathematical Background

In this section we would like to give some mathematical background from number theory which is used in the proposed scheme.

Two numbers $a$ and $b$ are said to be equal or congruent modulo $N$ iff $N|(a-b)$, i.e. iff their difference is exactly divisible by $N$. We write $a \equiv b (mod\ N)$.

The set of numbers congruent to $a\ modulo\ N$ is denoted by $[a]_N$. If $b \in [a]_N$ then, by definition $N|(a-b)$, or in other words, $a$ and $b$ have the same remainder of division by $N$. Since there are exactly $N$ possible remainders of division by $N$, there are exactly $N$ different sets $[a]_N$. Quite often these $N$ sets are simply identified with the corresponding remainders: $[0]_N = 0, [1]_N = 1, \ldots, [N-1]_N = N-1$. Remainders are often called residues; accordingly, $[a]'s$ are also known as the residue classes.

It's easy to see that if $a \equiv b\ (mod\ N)$ and $c \equiv d (mod\ N)$ then $(a+c) \equiv (b+d)(mod\ N)$.

The same is true for multiplication. These allow us to introduce an algebraic structure into the set $\{[a]_{[N]}: a = 0,1,\ldots N-1\}$.

By definition,
$$[a]_N + [b]_N = [a+b]_N$$

$$[a]_N * [b]_N = [a*b]_N$$

Subtraction is defined in an analogous manner.
$$[a]_N - [b]_N = [a-b]_N$$

It can be verified that the set $\{[a]_{[N]}: a = 0,1,\ldots N-1\}$ becomes a ring with commutative addition and multiplication. Division can't be always defined. The situation improves for prime $N$ 's in which case division can be defined uniquely. For prime $N$, the set $\{[a]_{[N]}: a = 0,1,\ldots N-1\}$ is promoted to a field. That is if $N$ is Prime, there exists inverse for every element in $Z_N$. The inverse of an element can be easily found out by using the extended Euclidean algorithm.

Let $a$ be one of the positive remainders of division by $P: 0 < a < P$ : $[a]_P, [2a]_P, [3a]_p, \ldots, [(P-1)a]_P$ are all different.. That is in the sequence $\{a, 2a, 3a, \ldots, (P-1)a\}$ no two numbers are congruent modulo $P$. Let there be two numbers $1 \le m < n < p$ such that $na \equiv ma\ (mod\ P)$. This would imply that $P|a(n-m)$. Since $P$ is prime either $P|a$ or $P|(n-m)$. But both $a$ and $(n-m)$ are positive integers less than $P$. So it can't divide either of them. So it leads to a contradiction.

It indeed follows that the set $\{[a]_P, [2a]_P, [3a]_p, \ldots, [(P-1)a]_P\}$ is just a permutation of the set $\{[1]_P, [2]_P, [3]_p, \ldots, [(P-1)]_P\}$.

If two sets are permutations of each other, then products of their elements are clearly equal.
$$[(P-1)!]_P = [1]_P \cdot [2]_P \cdot [3]_P \ldots [(P-1)]_P$$
$$= [a]_P \cdot [2a]_P \cdot [3a]_P \ldots [a(P-1)]_P$$
$$= [a^{P-1}(P-1)!]_P$$

Now, dividing by $[(P-1)!]_P$ gives $\quad 1 = [[a]^{P-1}]_P$ or
$$a^{P-1} = 1\ (mod\ P)$$

This is valid for any $a$ not divisible by $P$. The theorem is known as the Fermat's Little Theorem [1].

## 3. Proposed Schemes

### 3.1 (2, 2) Secret Sharing Scheme

Secret sharing algorithm consists of two phases Sharing Phase and Revealing phase. In the sharing phase shares correspond to the secret image is generated and shared between the participants. In the revealing phase secret image is reconstructed from the specified number of shares.

The scheme introduced is lossless and can be used with binary and gray scale images. The sharing and revealing phase of the secret sharing algorithm for a (2,2) scheme is given below.

### 3.1.1 Sharing Phase

1. Choose a $Z_P$ where $P=257$.

2. Let $s$ be the secret pixel to share $s \in \{0-255\}$.

3. Choose a random number $r$ (1-256) and corresponding inverse $r^{-1}$.

4. Create share1 pixel ($s1$).
   If $r = 256$ then $s1 = 0$ else $s1 = r$.

5. Create share2 pixel ($s2$).
   If $s = 0$ then $s = 256$
   $s2 = s * r^{-1} \ (mod \ P)$.
   If $s2 = 256$ then $s2 = 0$.

6. Repeat from step3 until all the pixels of the secret image are processed.

### 3.1.2 Revealing Phase

Multiply the pixel values at corresponding location from the two shares and get the secret value of the pixel.

1. If $s1 = 0$ then $s1 = 256$.
2. If $s2 = 0$ then $s2 = 256$.
3. $s = s1 * s2 \ (mod \ P)$.
4. If $s = 256$ then $s = 0$.
5. Repeating the above step for all the pixels in the shares will reveal the secret image.

### 3.2 (2, 3) Secret Sharing Scheme

The following is the proposed (2, 3) secret image sharing scheme. In the sharing phase three shares are generated and in the revealing phase any two of them can be used to retrieve the secret image. The share generated is of same size as the secret.

### 3.2.1 Sharing Phase

1. Choose a $Z_P$ where $P=257$.
2. Let $s$ be the secret to be shared corresponds to the specified pixel value (0-255).
3. If $s = 0$ then set $s = 256$. Find the corresponding cube root of $s$, and is "$a$".
4. Choose a random number "$r$" (1-256).
5. Share1 pixel value is generated by
   $s1 = r * a \ (mod \ P)$.
   If $s1 = 256$ then $s1 = 0$.

6. Share2 pixel value is generated by

$$s2 = r^2 * a \ (mod \ P)$$

If $s2 = 256$ then $s2 = 0$.

7. Share3 pixel value is generate by

$$s3 = r^4 * a \ (mod \ P).$$

If $s3 = 256$ then $s3 = 0$.

8. Repeat from step3 until all the pixels of the secret image are processed. This will generate three share images.

### 3.2.2 Revealing Phase

The secret image can be revealed from share1 and share2 by applying the following operation corresponds to each pixel in the two shares.

1. If $s1 = 0$ then $s1 = 256$.
2. If $s2 = 0$ then $s2 = 256$.
3. $a = (s1)^2 (s2)^{-1} (mod \ P)$
4. $s = a^3 (mod \ P)$
5. If $s = 256$ then $s = 0$.

The secret image can be revealed from share1 and share3 by applying the following operation corresponds to each pixel in the two shares.

1. If $s1 = 0$ then $s1 = 256$.
2. If $s3 = 0$ then $s3 = 256$.
3. $s = (s1)^4 \ (s3)^{-1} \ (mod \ P)$
4. If $s = 256$ then $s = 0$.

The secret image can be revealed from share2 and share3 by applying the following operation corresponds to each pixel in the two shares.

1. If $s2 = 0$ then $s2 = 256$.
2. If $s3 = 0$ then $s3 = 256$.
3. $a = (s2)^2 (s3)^{-1} \ (mod \ P)$
4. $s = a^3 \ (mod \ P)$
5. If $s = 256$ then $s = 0$.

## 4. Security Analysis

The probability of guessing a secret image pixel value is 1/256 except in the case of share2 in (2,3) scheme, which is 1/128. If the adversary is having a share then the probability of obtaining a secret image pixel is still 1/256. That is for an adversary, the probability of revealing a pixel value in the secret image is same when he is having a share or not. Hence the proposed scheme provides perfect security. Given a 256x256 shadow image the probability of obtaining the secret image is $(1/256)^{(256 \times 256)}$. The scheme is also ideal because the size of the share is same as secret size.

## 5. Experimental Results

The experimental results obtained for (2, 2) secret sharing scheme using the 256x256 gray scale image is shown in Fig.1. From the results it is clear that the proposed scheme is lossless and gave better results compared with the method used by the Chen and Chang. The computation involved is very less and hence it is more efficient compared with their scheme. The experimental results of the (2, 3) scheme is shown in Fig.2.

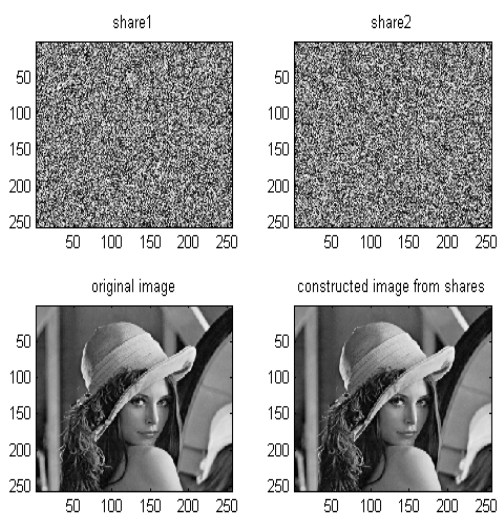

**Fig. 1 (2,2) secret image sharing using a 256x256 gray scale image**

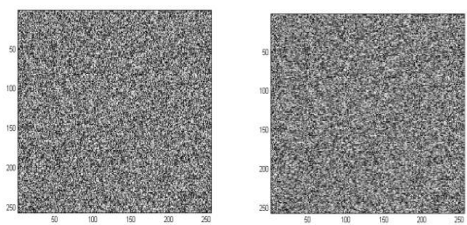

**(a )**          **(b)**

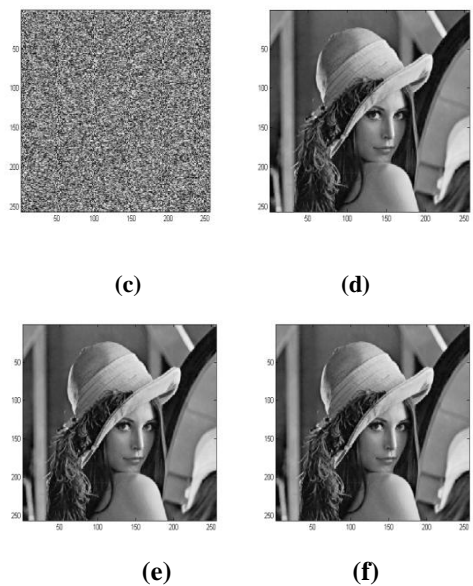

**(c)**          **(d)**

**(e)**          **(f)**

**Fig. 2 (2,3) secret image sharing using a 256x256 gray scale image.(a),(b)(c) are the 3 shares.(d) is the revealed secret image from (a) &(b). (e) is the revealed secret image from (a) & (c). (f) is the revealed image from (b) & (c).**

## 6.Conclusion

We have proposed a new method to share the secret image using basic number theoretic concept. The proposed (2,2) scheme is lossless and the computation involved is less complex than the (2, 2) scheme proposed by Chen and Chang using quadratic residues. Their scheme is lossy and in the lossless scheme the share size is larger. Both these problems are avoided in our scheme. The scheme can be extended to (n, n) scheme. The newly constructed (2, 3) scheme is also efficient compared with Shamir's scheme. The complexity involved in secret revealing of Shamir's scheme is overcome here with simple modular multiplication. The shares having the same size as the secret and hence the scheme is ideal. The scheme can be extended to the sharing of color images and text.